\documentclass[twocolumn,showkeys]{revtex4}
\usepackage{graphicx}
\usepackage{times}
\usepackage{amsmath}  % paquete matemático

\begin{document}

\title{Effects of AMM on the EoS of Magnetized Dense Systems}

\author{D. Manreza Paret}
\email{dmanreza@fisica.uh.cu}
\affiliation{Facultad de F{\'i}sica, Universidad de la Habana, San L{\'a}zaro y L, Vedado La Habana, 10400, Cuba}
\author{A. Perez Martinez}
\email{aurora@icimaf.cu}
\affiliation{Instituto de Cibern{\'e}tica Matem{\'a}tica y F{\'i}sica (ICIMAF) Calle E esq 15 No. 309 Vedado\\
La Habana, 10400, Cuba}
\author{E. J Ferrer}
\email{ejferrer@utep.edu}
\affiliation{Department of Physics, University of Texas at El Paso, El Paso, TX 79968, USA}
\author{V de la Incera}
\email{vincera@utep.edu}
\affiliation{Department of Physics, University of Texas at El Paso, El Paso, TX 79968, USA}

\date{\today}

\begin{abstract}
We investigate the effects of the anomalous magnetic moment (AMM) in the EoS of a fermion
system in the presence of a magnetic field. In the region of strong magnetic fields ($B>m^2$) the
AMM is found from the one-loop fermion self-energy. In contrast to the weak-field AMM found by
Schwinger, in the strong magnetic field case, the AMM depends on the Landau level (LL)
and decreases with it. The effects of the AMM in the EoS at intermediate-to-large fields can be found introducing the one-loop, LL-dependent AMM in the effective Lagrangian that is then used to find the thermodynamic potential of the system. We compare the plots of the parallel and perpendicular pressures versus the magnetic field in the strong field region considering the LL-dependent AMM, the Schwinger AMM, and no AMM at all. The results clearly show a separation between the physical magnitudes found using the Schwinger AMM and the LL-dependent AMM. This is an indication of the inconsistency of considering the Schwinger AMM beyond the weak field region $B< m^2$ where it was originally found. The curves for the EoS, pressures and magnetization at different fields give rise to the well-known de Haas van Alphen oscillations, associated to the change in the number of LL contributing at different fields.
\end{abstract}


\maketitle

\section{Introduction}

As it is known, the equation of state (EoS) plays a crucial role in determining the structure of neutron stars. The EoS is configured from the star inner content and external conditions as density, temperature, applied fields, etc. The fact that magnetic fields of strength sufficiently large to affect the matter state often exist in many compact objects, have motivated many studies of magnetic-field effects on stellar EoS's  \citep{Canuto, Chakrabarty1, Lattimer1, Aurora1, Chaichian, Richard1, Ferrer1, Paulucci, Chakrabarty2, Lattimer2, Aurora2, Dexheimer, Strickland}.

A magnetic field can affect a fermion system in various ways. It produces the well-known Landau quantization in the spectrum of the charged fermions.  It leads to an anisotropy between the parallel and transverse pressures. It can couple with the anomalous magnetic moment (AMM) of particles affecting their spectrum and consequently influencing the physical properties of the system. This coupling with the AMM can affect even neutral particles, as neutrons, which have an effective AMM due to their composite nature. The study of the effects produced by the AMM on magnetized neutron stars has unveil some interesting features \citep{Chakrabarty2, Lattimer2, Aurora2, Dexheimer, Strickland}, as for example, stiffening the EoS, a dramatic variation of the proton fraction that at very high magnetic fields will lead to pure neutron matter, etc. However, when considering potential AMM effects one should be careful of using values of the AMM that are consistent with the region of magnetic field strengths to be explored.

The AMM is commonly introduced via the so-called minimal coupling with the external field $\tau \sigma_{\mu\nu}F^{\mu\nu}$. For $\tau=\kappa\mu_B$, with $\kappa=\frac{\alpha}{2\pi}$ and $\mu_B=\frac{e}{2m}$, it reduces to the well-known, linear-in-field AMM found by Schwinger \citep{Schwinger1}. It is important however to keep in mind that Schwinger's  radiative correction to the magnetic moment is only valid at weak fields. Ignoring this limitation can lead to an overestimation of the AMM effects or, as pointed out many years ago by Jancovici \citep{Jancovici}, to wrong physical interpretations.

In this paper we calculate the one-loop radiative correction to the AMM in the strong field region ($m^2\ll B \le \mu^2$) and analyze its contribution to the main thermodynamical quantities and the EoS of a magnetized system of charged fermions. Just for comparison, and to underline the inconsistency of using Schwinger's AMM beyond the weak-field approximation, we include in our plots the results considering the AMM from Schwinger.

Even though the Maxwell contribution to the pressure and energy is important for applications to astrophysics, throughout the paper we ignore this contribution as we are interested here just in finding how the AMM affects the thermodynamical quantities and this only comes from the matter contribution to the thermodynamic potential.

The presentation is organized as follow. In Sec. \ref{section2} we extract the  AMM and mass contributions from the one-loop fermion self-energy in the presence of a constant and uniform magnetic field.  Both the AMM and the mass one-loop correction depend on the Landau level (LL) $l$. In Sec. \ref{section3}, we present the fermion's dispersion relations for an effective theory that incorporates the interaction of the external magnetic field with the AMM. We compare the dispersions considering the Schwinger AMM (beyond its region of validity) and the LL-dependent AMM found in Sec. \ref{section2}. In Sec. \ref{section4} the thermodynamical magnitudes of the system in the strong field region are obtained and compared with the ones found using the Schwinger approximation for the AMM and zero AMM. Finally, in \ref{section5}  we state our concluding remarks.

\section{AMM in the Strong Field Region}\label{section2}

Our goal is to extract the AMM from the one-loop fermion self energy $\Sigma^{l}(\overline{p})$ in a theory of charged fermions interacting with a constant, uniform and strong magnetic field ($B > m^2$). A constant and uniform magnetic field breaks the rotational symmetry of the system to the subgroup of rotations $O(2)$ about the field axis. This explicit symmetry breaking is reflected in the general structure of the fermion self energy, which, for a field arbitrarily taken along the third axis, can be written as \citep{Ferrer2}
\begin{equation}\label{SE-LLL}
\Sigma^{l}(\overline{p})
=Z_{\|}^{l}\overline{p}_{\|}^\mu\gamma_{\mu}^{\|}+Z_{\bot}^{l}\overline{p}_{\bot}^\mu\gamma_{\mu}^{\bot}+M^{l}I+iT^{l}\gamma^{1}\gamma^{2},
\end{equation}
with ${\overline{p}}_{\|}^\nu=(p^{0},0, 0,p^{3})$ the parallel and  ${\overline{p}}_{\bot}^\nu=(0,0, \sqrt{2eBl},0)$ with $l=0,1,2...$ the transverse momentum components; $Z_{\|}^{l}$ and $Z_{\bot}^{l}$ the wave function's renormalization coefficients;  and $M^{l}$ and $T^l$ the mass and anomalous magnetic moment that have to be determined as solutions of the Schwinger-Dyson (SD) equations of the theory in a chosen approximation.

For arbitrary Landau level  $l$, the one-loop self-energy is given by \citep{Ferrer3, Ferrer7}
{\setlength{\mathindent}{0pt}
\begin{multline}\label{SD-EqL}
\Sigma^{l}(\overline{p})\Pi(l)\hspace{-0.1cm}=
\hspace{-0.1cm}-ie^2(2eB)\Pi(l)\int\frac{d^4\widehat{q}}{(2\pi)^4}\frac{e^{-\widehat{q}^2_\bot}}{\widehat{q}^2}[L_l\\
+L_{l+1}+L_{l-1}], \quad l=0,1,2,\ldots
\end{multline}
where
{\setlength{\mathindent}{0pt}
\begin{eqnarray*}
 %\nonumber to remove numbering (before each equation)
  L_l &=& \gamma_{\mu}^{\|}G^{l}(\overline{p-q})\gamma_{\mu}^{\|}, \nonumber\\
  L_{l+1} &=& \Delta(+)\gamma_{\mu}^{\bot}G^{l+1}(\overline{p-q})\gamma_{\mu}^{\bot}\Delta(+), \nonumber\\
  L_{l-1} &=& \Delta(-)\gamma_{\mu}^{\bot}G^{l-1}(\overline{p-q})\gamma_{\mu}^{\bot}\Delta(-) \nonumber\\
  (\overline{p-q})_l&=&(p^{0}-q^{0},0,-\sqrt{2eBl},p^{3}-q^{3}) \nonumber
  \label{expresiones1}
\end{eqnarray*}
with
{\setlength{\mathindent}{0pt}
\begin{equation}
 %\nonumber to remove numbering (before each equation)
G^l(\overline{p})=\frac{\overline{p}\cdot\gamma+m}{\overline{p}^2-m^2}
  \label{FermionProp}
\end{equation}
the fermion propagator, and
{\setlength{\mathindent}{0pt}
\begin{equation}
\Delta(\pm) =\frac{I\pm i\gamma^{1}\gamma^{2}}{2},
\end{equation}
the spin projectors. The operator $\Pi(l)$,
{\setlength{\mathindent}{0pt}
\begin{equation}
 \Pi(l) =\Delta(\textrm{sgn}(eB))\delta^{l0}+I(1-\delta^{l0}),
\end{equation}
appears because only one spin projection contributes to the lowest Landau level (LLL). Here we are assuming, without lost of generality, $B$ along z and a positively charged fermion.
Normalized quantities, denoted with a hat, are defined by $\widehat{q}_\mu=q_\mu/\sqrt{2eB}$.  To find the expression (\ref{SD-EqL}) for the self-energy, we summed in all the internal Landau levels and considered a strong-field approximation.

Taking into account the algebra of the spin projectors $\Delta(\pm)\Delta(\pm)=\Delta(\pm)$, $\Delta(\pm)\Delta(\mp)=0$,  one  can readily extract the LLL ($l=0$) contribution to Eq. (\ref{SD-EqL}), which after Wick rotation ($p^{0}\rightarrow -ip^4$), becomes
{\setlength{\mathindent}{0pt}
\begin{multline}\label{LLLMT}
(M^0(p)+T^0(p))\Delta(+)=e^2(2eB)\int\frac{d^4\widehat{q}}{(2\pi)^4}
\frac{e^{-\widehat{q}^2_\bot}}{\widehat{q}^2}\\
\times\left(\frac{2m}{(\overline{p-q})^2_0+m^2} +\frac{2m}{(\overline{p-q})^2_{1}+m^2}\right )\Delta(+),
\end{multline}

Note  that Eq.\,(\ref{LLLMT}) reflects  the fact that fermions in the LLL have only one spin orientation, and as a consequence, it is impossible to determine $M^0$ and $T^0$ independently  \citep{ Ferrer3, Ferrer7}.
For $l\neq 0$ we obtain
\begin{multline}\label{LLnq0}
T^{l} = \frac{e^2m}{16\pi^2}\int d\widehat{q}_{\bot}^2e^{-\widehat{q}^2_\bot}  \left[\frac{\ln\frac{\widehat{q}_{\bot}^2}{\widehat{m}^2+l+1}}{\widehat{q}_{\bot}^2-\widehat{m}^2-l-1}\right.\\
\left.-\frac{\ln\frac{\widehat{q}_{\bot}^2}{\widehat{m}^2+l-1}}{\widehat{q}_{\bot}^2-\widehat{m}^2-l+1} \right],
\end{multline}

\begin{multline}\label{LLnq0M}
M^{l} = \frac{e^2m}{16\pi^2}\int d\widehat{q}_{\bot}^2e^{-\widehat{q}^2_\bot}
\left[\frac{2\ln\frac{\widehat{q}_{\bot}^2}{\widehat{m}^2+l}}{\widehat{q}_{\bot}^2-\widehat{m}^2-l}\right.\\
\left.+\frac{\ln\frac{\widehat{q}_{\bot}^2}{\widehat{m}^2+l+1}}{\widehat{q}_{\bot}^2-\widehat{m}^2-l-1}
+\frac{\ln\frac{\widehat{q}_{\bot}^2}{\widehat{m}^2+l-1}}{\widehat{q}_{\bot}^2-\widehat{m}^2-l+1} \right].
\end{multline}

In Eqs. (\ref{LLnq0}) and (\ref{LLnq0M})  we already took the infra-red limit ($p_0=0, p_3\rightarrow 0$) and integrated in $\widehat{q}_{\|}^2$. Following the same steps in (\ref{LLLMT}) we find

{\setlength{\mathindent}{0pt}
\begin{multline}\label{LLLMT2}
M^{0}+T^{0} = \frac{e^2m}{8\pi^2}\int d\widehat{q}_{\bot}^2e^{-\widehat{q}^2_\bot}  \left[\frac{\ln\frac{\widehat{q}_{\bot}^2}{\widehat{m}^2}}{\widehat{q}_{\bot}^2-\widehat{m}^2}\right.\\
\left.+\frac{\ln\frac{\widehat{q}_{\bot}^2}{\widehat{m}^2+1}}{\widehat{q}_{\bot}^2-\widehat{m}^2-1} \right],
\end{multline}
which in the leading strong-field approximation reduces to
\begin{equation}\label{LLLMT32}
    E^0=M^{0}+T^{0} =\frac{\alpha}{4\pi}m\textrm{log}^2(\widehat{m}^2).
\end{equation}

Fig.\,\ref{fig0} shows the one-loop contribution of the AMM to the fermion self-energy versus $2eB/m^2$ for $l=0,1,5$ and for Schwinger. Notice that just by increasing the LL in a few units, the AMM becomes negligible.  This is a consequence of the different sign between the two terms in  (\ref{LLnq0}), which tend to cancel out with increasing $l$. In contrast, the Schwinger contribution $T_{Sch}= \kappa\mu_B B$, which is linear in the field, increases significantly, but this occurs in a region where the validity of the linear approximation used to find this AMM is not correct.
\begin{figure}[h!]
\includegraphics[width=80mm,height=50mm]{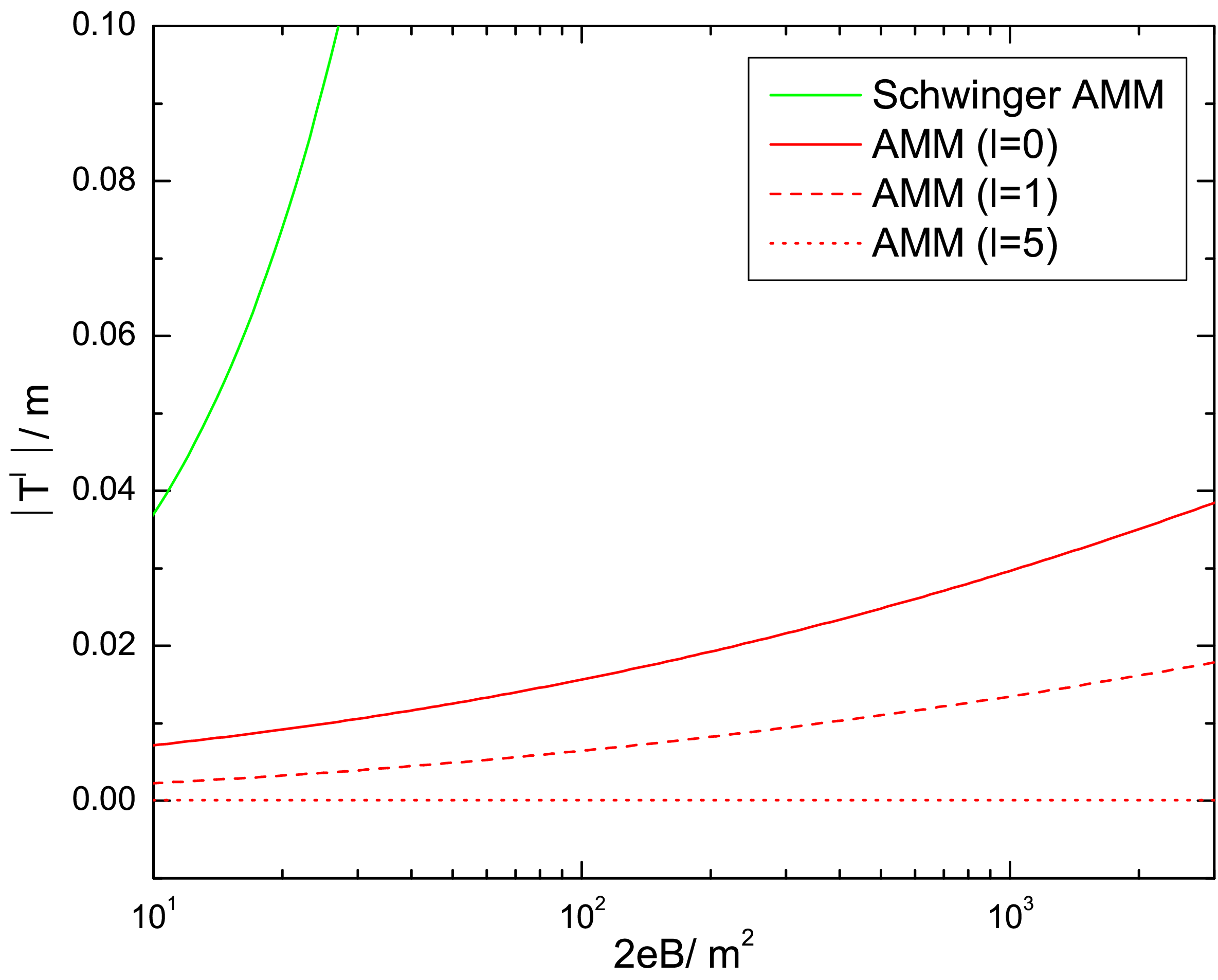}
\caption{Variation of the AMM contribution to the self-energy with the field for different Landau levels. Notice that the AMM becomes negligible just after a few levels. Considering the Schwinger AMM, which is the same for all the levels, completely overestimates the actual physical AMM value in the strong-field region.} \label{fig0}
\end{figure}
\section{Dispersion Relations}\label{section3}

We can incorporate the vacuum quantum effects in the fermion propagator by introducing the corrections coming from the one-loop fermion self-energy, that is, the  AMM and the radiative correction to the mass.  With these quantum corrections, the dispersion relations of the fermions are found from
{\setlength{\mathindent}{0pt}
\begin{equation}
detG^{-1}_l=det[\overline{p}\cdot\gamma-m-M^lI-iT^l\gamma^1\gamma^2]=0,
\end{equation}\label{Inverse-Prop}
For $l\geq 1$ they take the form
{\setlength{\mathindent}{0pt}
\begin{multline}\label{isp-Rel-1}
\epsilon_{\sigma, l}^2=p_3^2+[\sqrt{(m +M^l)^2 +2|eB|l}+\sigma T^l)]^2, \\
\, l\geq 1, \, \sigma=\pm 1
\end{multline}
while for the LLL ($l=0$) they become
{\setlength{\mathindent}{0pt}
\begin{equation}\label{Disp-Rel-3}
\epsilon_{1,0}^2=p_3^2+(m+M^0+T^0)^2
\end{equation}

In the weak-field case studied by Schwinger \citep{Schwinger1}, the dispersion relation becomes
{\setlength{\mathindent}{0pt}
\begin{equation}
detG^{-1}_l=det[\overline{p}\cdot\gamma-m-\kappa\mu_BB\sigma]=0,
\end{equation}\label{Inverse-Prop1}
and the energy spectrum is
{\setlength{\mathindent}{0pt}
\begin{equation}
 \epsilon_{\sigma,l}^2=p_3^2 + [(m^2+2eBl)^{1/2}-\kappa\mu_BB\sigma]^2  \label{Sch}
\end{equation}
Notice that the dispersion relations based on Schwinger  weak-field results do not include the radiative correction to the mass. They have been been used in this way in many works in the literature, and here we keep the same  approach in order to compare with the literature's results and also to underline where it may not be appropriate. For all the $l>0$, ignoring the quantum correction to the mass is actually fine, as this correction just enters as an addition to the fermion mass, and it is negligible compared to the bare fermion mass.

A problem may appear however if one follows the same approach with the LLL dispersion.  As the Schwinger AMM is independent of the LL, in the past some authors  \citep{O'Connell, Canuto2, Canuto3} considered that the lowest rest energy of the fermions, which would correspond to the LLL, should be
{\setlength{\mathindent}{0pt}
\begin{equation}\label{energy1}
\varepsilon^{0}_{\text{\tiny Schw}}=\mid m -\kappa\mu_B B\mid,
\end{equation}
We do not agree that this is the correct expression of the LLL rest energy in the Schwinger approximation, even if the field is kept  in the weak-field approximation where Schwinger AMM is valid.  As can be seen from (\ref{SE-LLL}), in the LLL the single spin projection of the fermions makes it impossible to find the AMM separately from the mass, so in this case both the mass and the AMM corrections enter as a net correction to the bare mass.
 This statement is independent of the field-approximation considered. We will return to a detailed discussion of this issue in a future paper \cite{Ferrer6}. Since (\ref{energy1}) has been considered in the literature as the LLL particle rest energy and just for the sake of understanding, we will compare $\varepsilon^{0}_{\text{\tiny Schw}}$  with the LLL rest energy found within our strong field approximation
{\setlength{\mathindent}{0pt}
\begin{equation}\label{Disp-Rel-4}
\varepsilon^{0}_{LLL} =\mid m+E^0\mid,
\end{equation}
to illustrate how much one may over- or under-estimate the physical quantities when using the Schwinger AMM beyond the region of validity of the linear approximation.

 In Fig.\,\ref{fig1} we plot  $\varepsilon^0_{LLL}$ and $\varepsilon^{0}_{\text{\tiny Schw}}$ versus $2eB/m^2$ for strong field values. As expected, the behaviors of $\varepsilon^{0}_{\text{\tiny Schw}}$ and $\varepsilon^0_{LLL}$ are very different. If one mistakenly calculates the thermodynamical quantities, using $\varepsilon^{0}_{\text{\tiny Schw}}$ for the LLL rest energy of fermions, all kind of wrong effects may arise. The problem becomes more relevant with increasing magnetic field. At strong fields, $\varepsilon^{0}_{\text{\tiny Schw}}$  deviates completely from the correct LLL rest energy, as this figure shows, so using it may lead to totally incorrect effects.

 On the other hand, for the weak-field region, the influence of $\varepsilon^{0}_{\text{\tiny Schw}}$ in the physical magnitudes becomes more and more significant when the field approaches $m^2$, a region where the validity of the weak-field approximation becomes rather questionable. On the other hand, we suspect that the use of (\ref{energy1}) in the thermodynamic potential may contribute to the reported stiffening of the EoS produced by the AMM  \citep{Lattimer1, Strickland}.  One should be careful then when doing astrophysical interpretations of the AMM effects if these two issues are not properly addressed.

\begin{figure}[ht!]
\begin{center}
\includegraphics[width=80mm,height=50mm]{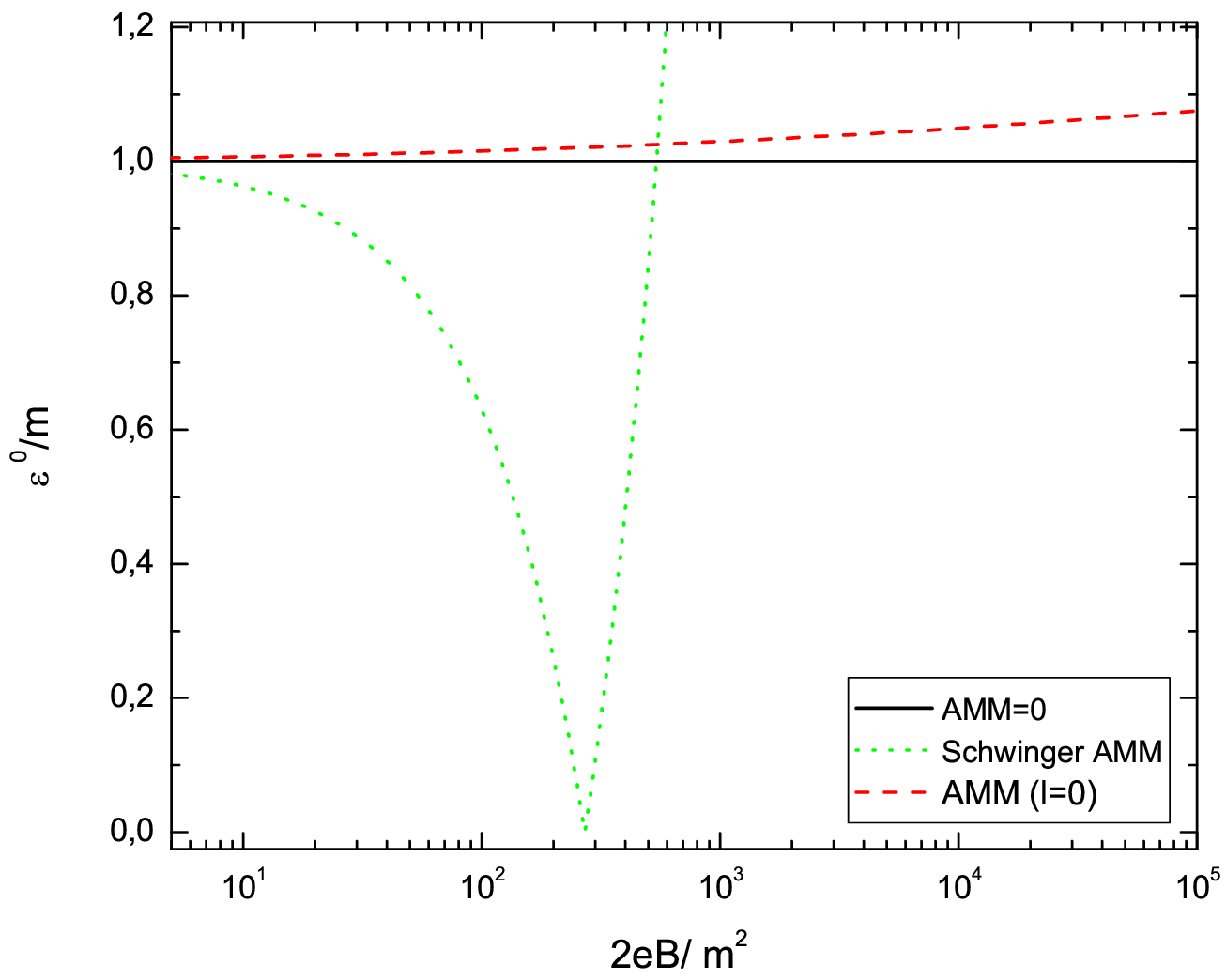}
\caption{$\varepsilon^0$ and $\varepsilon^{0}_{\text{\tiny Schw}}$ versus $2eB/m^2$ for strong field values} \label{fig1}
\end{center}
\end{figure}

Notice that the threshold field marking the validity of the Schwinger AMM approximation is $B\simeq 10^{13}$ G for electrons and $B\simeq 10^{15}$ G for u quarks with current mass of $5 MeV$.

\section{Thermodynamical Magnitudes}\label{section4}

Let us find the impact of the AMM in the thermodynamical magnitudes as particle density, energy density, magnetization and parallel and perpendicular pressures. With this goal in mind, we shall find the matter contribution $\Omega_M$  to the thermodynamical potential of the theory, including the one-loop radiative contributions found in Sec. \ref{section2} (and Schwinger again just for comparison) and considering the system at finite temperature and density

{\setlength{\mathindent}{0pt}
\begin{multline}\label{Grand-Potential-4}
\Omega_M= -\frac{eB}{\beta}\left[\sum_{p_4}\int\limits_{-\infty}^{\infty}\frac{dp_3}{(2\pi)^2} \ln \det G^{-1}_0(\overline{p}^{*})\right.\\
\left.+ \sum_{\sigma\pm1}\sum_{l=1}^{\infty}\sum_{p_4}\int\limits_{-\infty}^{\infty}\frac{dp_3}{(2\pi)^2} \ln \det G^{-1}_l(\overline{p}^{*})\right],
\end{multline}
here ${\overline{p}}^{*}_\nu=(ip^{4}-\mu,0, \sqrt{2eBl},p^{3})$ and $\beta$ is the inverse absolute temperature.

The thermodynamic potential $\Omega_M(B,\mu,T)$  can be split into a vacuum contribution $\Omega^V_M(B,0,0)$, which needs to be renormalized and depends on the magnetic field, and a statistical contribution $\Omega^S_M(B,\mu,T)$,
\begin{equation}\label{Omega1}
  \Omega_M= \Omega^V_M(B,0,0) + \Omega^S_M(B,\mu,T).
\end{equation}
The renormalized vacuum contribution was found in \citep{Ferrer5}.
In the following derivations we neglect it, since we are interested here in a region of fields $2eB< \mu^2$ where the thermodynamic contribution to the physical magnitudes is more important than the vacuum one.

As mentioned in the Introduction, we are not including the Maxwell term, since in this paper we are only interested in reporting how the thermodynamical magnitudes depend on the AMM. However, we call attention that for any application to neutron stars, the Maxwell contribution should not be ignored.

We take the zero temperature limit of Eq.\,(\ref{Grand-Potential-4}), which is of more interest for eventual applications to neutron stars, where $\mu\gg T$. The  zero-temperature, statistical part then becomes
%poner referencia
{\setlength{\mathindent}{0pt}
\begin{multline}\label{Om_mag}
\Omega^{S0}_{M} \equiv \Omega^S_M(B,\mu,0)\hspace{-0.1cm} =\hspace{-0.1cm}-m\frac{eB}{4\pi ^2}\left[\Omega_{LLL}\right. \\
\left.+ \sum_{\sigma=\pm 1} \sum_{l=1}^{l_{max}} \left( \mu\,{p}_{F}^{\sigma} -  ({\varepsilon}_{\sigma,l}^{0})^2\ln\frac{ {\mu}
+ {p}^{\sigma}_{F}}{{\varepsilon}_{\sigma,l}^{0}}\right)\right],
\end{multline}
where
{\setlength{\mathindent}{0pt}
\begin{eqnarray}\label{lmax}
l_{max}=\begin{cases}
I[\frac{(\mu-\sigma T^l)^2-(m+M^l)^2}{2eB}], & \text{AMM},\vspace*{2.0mm}\\
I[\frac{(\mu-\sigma\kappa\mu_BB)^2-m^2}{2eB}], & \text{Schwinger} \,\, \text{AMM}.
\end{cases}
\end{eqnarray}
where $I[z]$ denotes the integer part of $z$.

In Eq.\,(\ref{Om_mag}),
{\setlength{\mathindent}{0pt}
\begin{equation}\label{Om_mag0}
   {\Omega}_{LLL}\hspace{-0.1cm} =\hspace{-0.1cm}\left ({\mu}\,{p}_{F}^{0} -({\varepsilon}_{1,0}^{0})^2
\ln\frac{  {\mu} +{p}_{F}^0}{{\varepsilon}_{1,0}^0}\right ),
\end{equation}
is the LLL contribution, and
{\setlength{\mathindent}{0pt}
\begin{eqnarray} \label{dimvar}
   {p}_{F}^{\sigma}\hspace{-0.1cm}=\hspace{-0.1cm}\sqrt{ {\mu}^2-({\varepsilon}_{\sigma,l}^{0})^2}, \,\, \quad {p}_F^{0}=\sqrt{{\mu}^2-({\varepsilon}_{1,0}^{0})^2},
\end{eqnarray}
{\setlength{\mathindent}{0pt}
\begin{eqnarray}\label{restenergy21}
{\varepsilon}_{\sigma,l}^{0}=\begin{cases}
|\sqrt{(m +{M}^l)^2 +2|e{B}|l}+\sigma {T}^l|, & \text{AMM},\vspace*{2.1mm}\\
|\sqrt{m^2+2e{B}l}-\sigma\kappa\mu_B{B}|,     & \text{Schwinger}.
\end{cases}
\end{eqnarray}

Since we are approximating $\Omega_M \approx \Omega^{S0}_{M} $,  the zero temperature matter contribution to the magnetization $\cal {M}$, particle density $N$, energy density, and pressures can be found from the following formulas
{\setlength{\mathindent}{0pt}
\begin{eqnarray}\label{Magnetizacion}
\mathcal{M}\hspace{-0.1cm}=\hspace{-0.1cm}-(\partial\Omega^{S0}_{M}/\partial B)=\frac{e m}{4\pi ^2}\left\{  \mathcal{M}_{LLL} + \sum_{l=1}^{l_{max}}\sum_{\sigma=\pm 1} {\mu}{p}_F^{\sigma}\right.\nonumber\\
\left. -\left[ ({\varepsilon}_{\sigma,l}^{0})^2 + 2{\varepsilon}_{\sigma,l}^{0} {C}^l \right]
       \ln\frac{{\mu}+ {p}^{\sigma}_F}{{\varepsilon}_{\sigma,l}^{0}}\right\},
\end{eqnarray}
{\setlength{\mathindent}{0pt}
\begin{multline}\label{Density1}
N=-(\partial\Omega^{S0}_{M}/\partial \mu)= \\
\frac{m^2}{2\pi^2}\frac{B}{B^c}\left( {p}_F^0+\sum_{l=1}^{l_{max}}\sum_{\sigma=\pm 1}{p}_F^{\sigma}\right),
\end{multline}
{\setlength{\mathindent}{0pt}
\begin{eqnarray}\label{EnerPresure1}
   E&=& \Omega^{S0}_{M} + \mu N,\nonumber\\
   P_\parallel  &=&-\Omega^{S0}_{M},\nonumber\\
   P_\bot &=&-\Omega^{S0}_{M}-B\mathcal{M},
\end{eqnarray}
where
{\setlength{\mathindent}{0pt}
\begin{eqnarray}
  \mathcal{M}_{LLL}\hspace{-0.1cm}=\hspace{-0.1cm} {\mu}{p}_F^{0} -\left[ ({\varepsilon}_{1,0}^{0})^2+2{\varepsilon}_{1,0}^{0} {C}^{0} \right]\ln\frac{{\mu}+ {p}^{0}_F}{{\varepsilon}_{1,0}^{(0)}}, \\
  {C}^l \hspace{-0.1cm}=\hspace{-0.1cm} \begin{cases}
\frac{\frac{B}{B^c}l}{\sqrt{2l\frac{B}{B^{c}}+1}}+\sigma B\frac{\partial {T}^l(B) }{\partial B}, & \text{AMM}, \vspace*{2.0mm}\\
\frac{\frac{B}{B^c}l}{\sqrt{2l\frac{B}{B^{c}}+1}}-\sigma\kappa\mu_BB, & \text{Schwinger},
\end{cases}\\
\vspace*{4.0mm}
{C}^0\hspace{-0.1cm} =\hspace{-0.1cm} \begin{cases}
 B\frac{\partial \left[M^0(B)+T^0(B)\right]}{\partial B} & \text{AMM}, \vspace*{2.0mm}\\
 -\kappa\mu_BB, & \text{Schwinger}.
\end{cases}
\end{eqnarray}
where we have introduced $B^c=m^2/e$.
\section{Numerical Results}\label{section5}

For the numerical results we  concentrate in the  region of fields  $0.1<2eB/\mu^2<1$, which is consistent with all the assumptions made in our calculations.
In Fig.\,\ref{fig2}, Fig.\,\ref{fig3} and Fig.\,\ref{fig4} we plotted the normalized quantities $\mathcal{M}\cdot B/\mu^4$,  $E/\mu^4$, and $P_\|/\mu^4$, $P_\bot/\mu^4$ respectively, versus $2eB/\mu^2$ for three cases: one containing the radiative contribution of the LL-dependent AMM  (Eqs.\,(\ref{LLnq0})-(\ref{LLLMT2})),  other with the Schwinger AMM, and another without AMM.

In all the graphs we can see that as the magnetic field grows, the plots using the Schwinger AMM differ more and more from those with the LL-dependent AMM or with zero-AMM.
The last two practically overlap in the entire region, indicating that for strong field, the AMM is not relevant for the thermodynamical properties of the system.
\begin{figure}[ht!]
\begin{center}
\includegraphics[width=80mm,height=50mm]{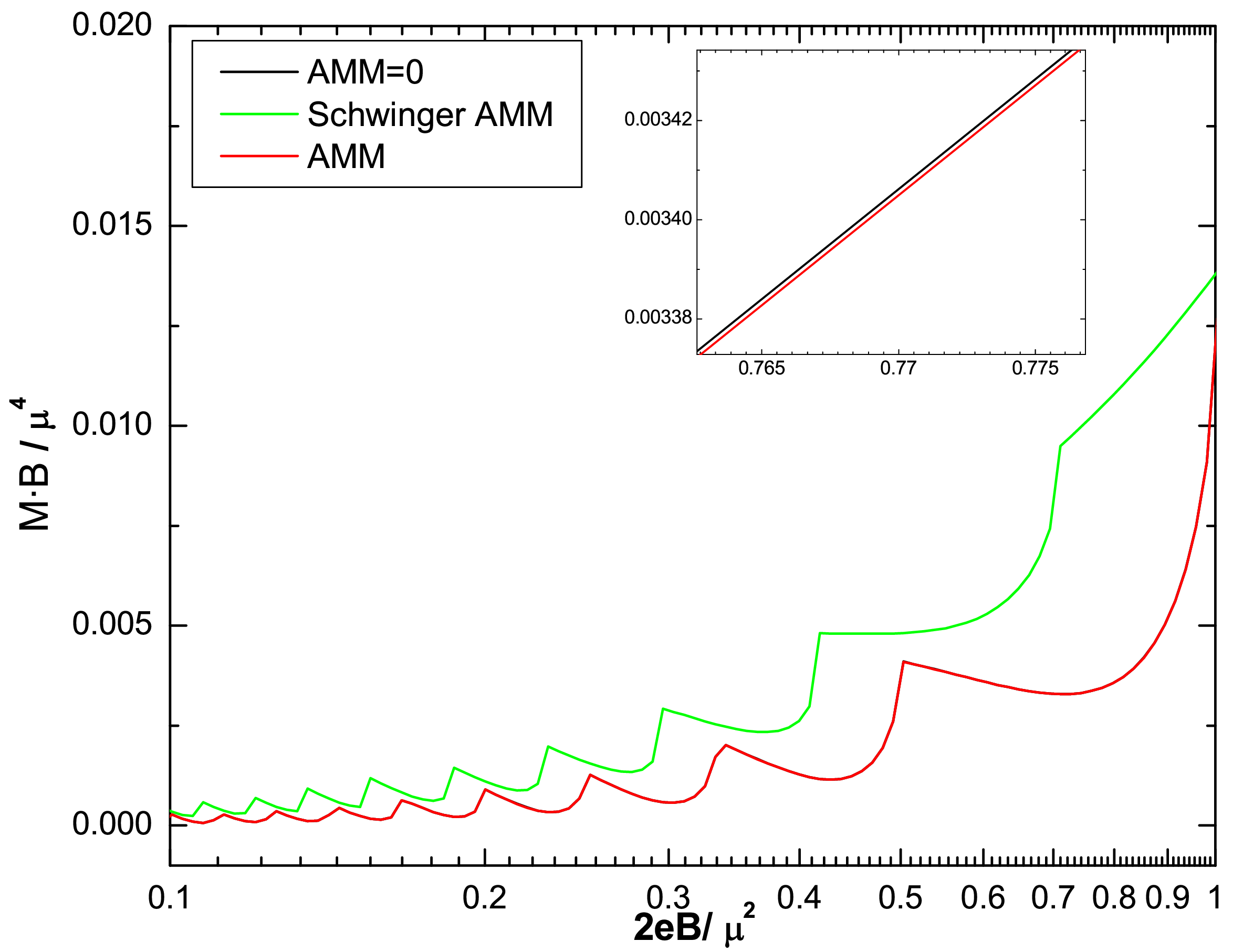}
\caption{Magnetization versus $2eB/\mu^2$. Comparison between AMM, Schwinger approximation of AMM, and  without AMM cases.} \label{fig2}
\end{center}
\end{figure}

With the exception of the magnetization, using the Schwinger AMM underestimates the corresponding physical magnitude. In all the cases, using the Schwinger AMM leads to an overestimation of the relevance of the particle's magnetic moment in the thermodynamics of the system. When one uses the consistent approximation for the AMM in the strong-field region, the effect of the AMM in the EoS is practicably negligible. This is in sharp contrast with the claims made at weak fields \citep{Lattimer1, Strickland}, where the AMM from Schwinger has been interpreted as responsible for the stiffness of the EoS. It is hard to understand how the magnetic moment can contribute to make the EoS stiffer at weak fields and do nothing at much stronger fields. This is another reason that  points to some potential issues with the use of $\varepsilon^{0}_{\text{\tiny Schw}}$  in the LLL dispersion, as already pointed out.

\begin{figure}[ht!]
\begin{center}
\includegraphics[width=80mm,height=60mm]{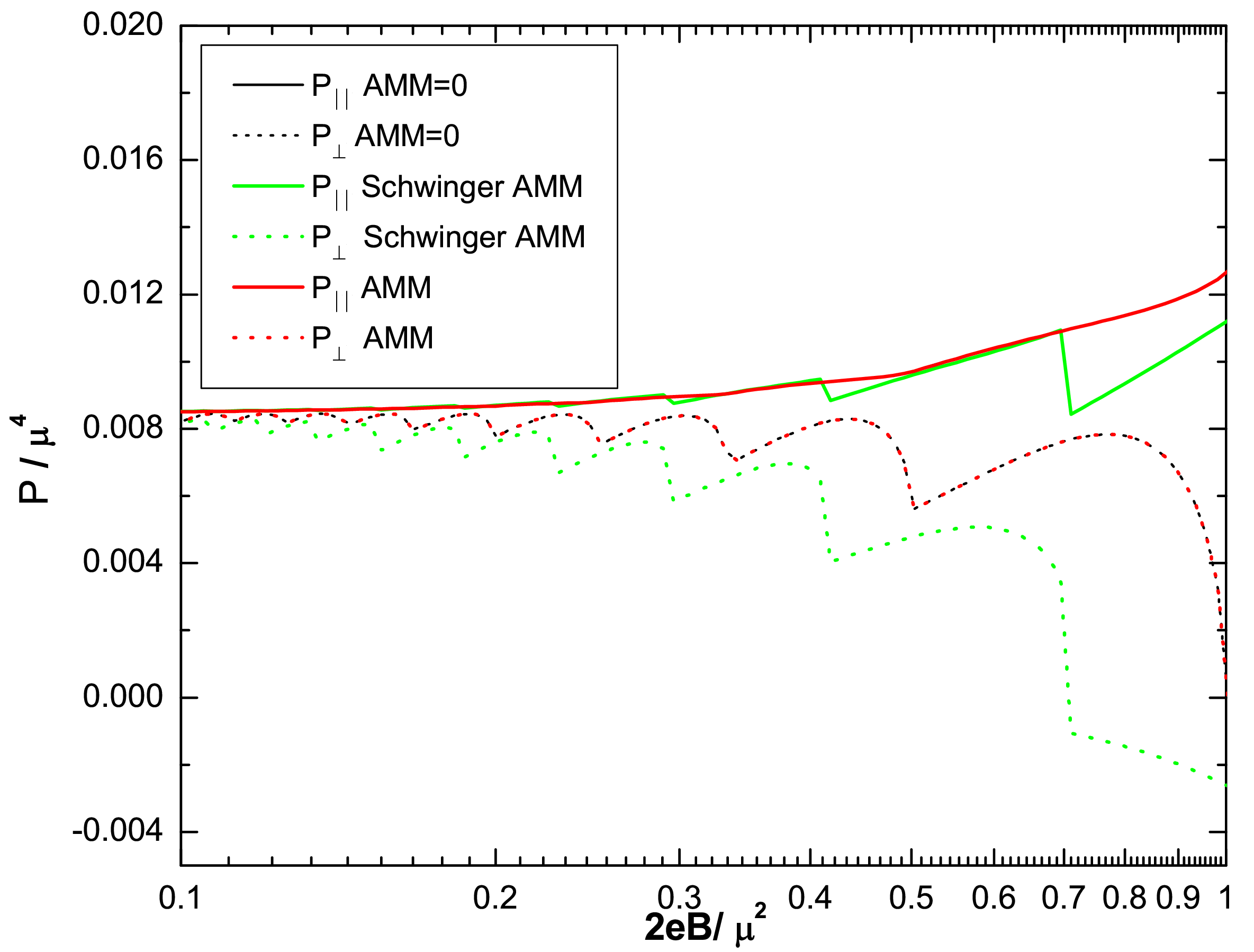}
\caption{Pressures parallel and perpendicular versus $2eB/\mu^2$, with $\mu=300 MeV$. Comparison between AMM, Schwinger approximation of AMM, and  without AMM cases. } \label{fig3}
\end{center}
\end{figure}

\begin{figure}[ht!]
\begin{center}
\includegraphics[width=80mm,height=50mm]{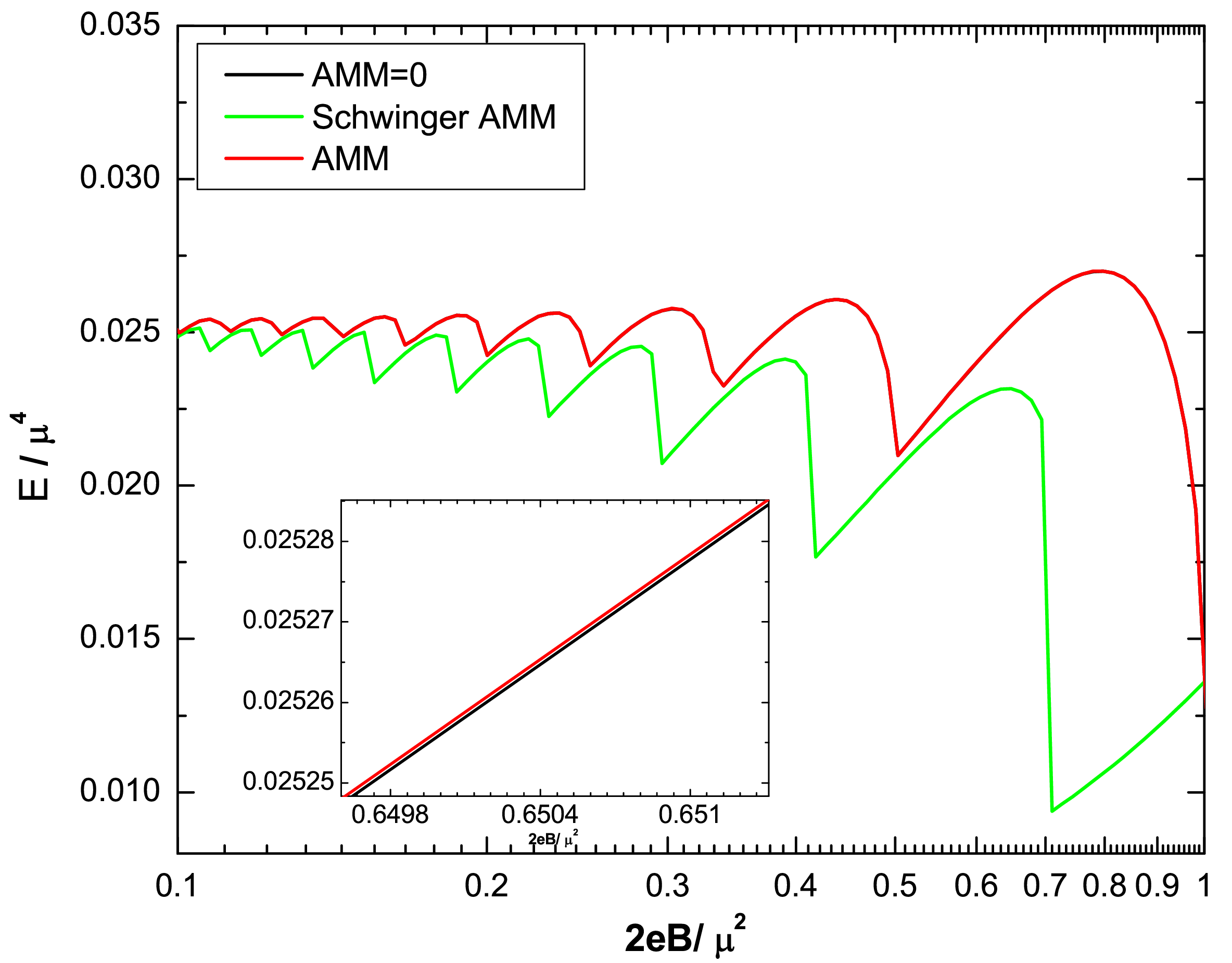}
\caption{Energy density versus $2eB/\mu^2$, with  $\mu=300 MeV$.. Comparison between AMM, Schwinger approximation of AMM, and  without AMM cases. } \label{fig4}
\end{center}
\end{figure}

\section{Conclusions}
In this paper we calculate the one-loop radiative correction of the AMM in the strong field region ($m^2\ll B \le \mu^2$) and analyze its contribution to the main thermodynamical quantities and to the EoS of a magnetized system of charged fermions. Just for comparison, and to underline the inconsistency of using Schwinger's AMM beyond the weak-field approximation, we include in our plots the results considering the AMM from Schwinger.

Our main finding is that the AMM has a negligible effect on the thermodynamical magnitudes of the system in the strong-field region. We did not observe any stiffness of the EoS, not even for the strongest fields considered. Our findings are at odd with several previous literature results \citep{Chakrabarty2, Lattimer2, Aurora2, Dexheimer, Strickland, Mao}. On one hand, we underline that considering the Schwinger AMM correction in the fermion propagator is only consistent at very weak fields. Hence any conclusion reached by using the Schwinger AMM at fields beyond $m^2$ is simply incorrect. Moreover, issues with the weak-field approximation may even show up already when the field approaches the critical value $eB^c=m^2$, from below, so any conclusion based on the behavior of the physical magnitudes in this region should be taken with a grain of salt. Finally, considering only the AMM, but ignoring the radiative mass correction in the LLL dispersion relation is inconsistent, because these two magnitudes are of the same order and enter as a shift to the bare mass.  Given that the stiffness of the EoS has been found precisely in the region close to the threshold field, where both the weak-field approximation becomes less effective and the LLL becomes more relevant, we suspect that they both are the culprit behind the odd effect of the AMM in the EoS.

We admit that at this point all our claims about the issues with the effects of the AMM at weak fields are just a hunch supported by some hand-waiving arguments. In the near future, we plan to do the calculations of the EoS in the weak-field region using the Schwinger AMM and considering the correct radiative contribution to the LLL dispersion relation to corroborate or disprove our claims.

\bigskip
A.P.M and D.M.P A.P.M. thanks to A. Cabo and H. Perez Rojas for discussions. The work of EJF and VI was partially supported by Nuclear Theory DOE grant DE-SC0002179. The work of A.P.M and D.M.P. have been supported  under the grant CB0407 and the ICTP Office of External Activities through NET-35.

%\bibliographystyle{mibib}
%\bibliography{smfns}

\end{document}